\documentclass[a4paper,twoside,10pt]{article}
\usepackage{graphicx,publaob}

\def\papertype{Contributed paper}
\begin{document}

\title{TESTING THE PERFORMANCE OF THE MILANKOVI\'C TELESCOPE}

\authors{A. Vudragovi\' c$^1$, M. B\'ilek$^2$, O. M\"uller$^2$, S. Samurovi\'c$^1$ \lowercase{and} M. Jovanovi\'c$^1$}

\address{$^1$Astronomical Observatory, Volgina 7, 11060 Belgrade, Serbia}
\Email{ana}{aob}{rs}
\address{$^2$Université de Strasbourg, CNRS, Observatoire Astronomique de Strasbourg (ObAS), UMR 7550, 67000 Strasbourg, France}

\markboth{TESTING PERFORMANCE OF THE MILANKOVI\'C TELESCOPE}


\abstract{We have undertaken a multi-band imaging campaign of several galaxies to study their low surface brightness features such as shells and streams. Using the 1.4-m Milankovi\'c telescope, we measured the surface brightness limits in various bands depending on the exposure time. Remarkably, within three to four hours of observations with the $L$-filter we reached the surface brightness limit of $\mu_g$=28.5-29.0 mag/arcsec$^2$. 
We have confirmed the faint stream of the elliptical galaxy NGC\,474 discovered with MegaCam.
The comparison to other deep photometric surveys has revealed that within few hours of observations we can produce competitive results, showing that the Milankovi\'c telescope can be a valuable asset in the study of the low-surface brightness Universe.}

\section{INTRODUCTION}

The low surface brightness Universe received renewed attention in the last decade. Several specially designed and envisioned projects started to collect very deep data (e.g., van Dokkum et al. 2014; Duc et al. 2015; Rich et al. 2020; B\'ilek et al. 2020). Deep imaging revealed low surface brightness features in the form of stellar streams and shells, and extended optical disks (almost the size of the HI disks) around nearby galaxies. 

According to the $\Lambda$CDM paradigm, massive galaxies formed hierarchically by consuming surrounding smaller galaxies (Frenk \& White 2012). The accretion of dwarf satellites had left stellar streams in the outskirts of their massive hosts that are long lived and memorize past events. Cosmological simulations predict numerous disruption signatures left over from the formation and evolution of massive galaxies (Bullock et al. 2005; Cooper et al. 2010). The detection of such features and, moreover, subsequent analysis involving their quantification, provides a strong test for this hierarchical model of galaxy formation. However, these features are hidden in the low-surface brightness regime well beyond 26 mag/arcsec$^2$.

Reaching the depth required to study low surface brightness structures, which are around 30 mag/arcsec$^2$ requires special observation technique and careful background subtraction. The method widely used is a technique called dithering -- off-setting the telescope randomly to sample different portions of the sky. This dithering pattern has to be large enough to make the target galaxy fall on different parts of the CCD chip. The result is  a correction that can be used to remove the prevalent systematics in the image, such as residuals from the flat-fielding or the sky background.

The quantification of the observation depth that allows comparison between different surveys is not well defined. There are in principle two ways to measure the surface brightness limit reached by an observation: (1) finding the limiting magnitude of the surface brightness profile of the object itself; or (2) quantifying background fluctuations. The problem with the former is that extrapolations of the profile can easily lead to misinterpretations of the true depth reached. The latter is independent of the modelling of the target galaxy, allowing for a better comparison between observations, which is why we use this approach. 

To study the capabilities of the Milankovi\'c telescope mounted at the Astronomical Station Vidojevica near Prokuplje (Serbia) and equipped with the IkonL CCD camera, we have imaged two galaxies for several hours -- namely NGC~474 and NGC~467 -- in four filters ($B$, $V$, $I$ and $L$). 
In Section 2 of this article, we describe the data acquisition and data reduction. In Section 3 we compare the surface brightness limit reached here to other surveys like the Sloan Digital Sky Survey (SDSS; York et al. 2000), the Beijing Arizona Sky Survey (BASS; Zou et al. 2017) and the Dark Energy Survey (DES; Dey et al. 2019). And finally, in Section 4 we discuss the results.

\section{OBSERVATIONS AND DATA REDUCTION}

\subsection{NGC~474 galaxy: $L$-band}

We have imaged the elliptical galaxy NGC~474 in the $L$-band on  October 22$^{\rm nd}$ 2019 with the 1.4m Milankovi\'c telescope using the IkonL CCD camera.
We have applied a randomized dithering pattern with offsets of $\sim$300 pixels ($\sim$2 arcsin). The individual exposure times were 300\,s and the airmass ranged from 1.29 to 1.92. We have taken 33 exposures using the $L$-filter, resulting in an  on-source integral exposure time of 2.75\,h. 
The image quality was on average 1.3 arcsec. An individual frame covers a square box of 13.3 arcmin, utilizing the focal reducer. As a result of the dithering the final co-add has the dimensions of 17.5 $\times$ 17.9 arcmin$^2$ (Figure 1, left). Previously, NGC~474 was observed by Duc et al. (2015) using the 3.6m Canada–France–Hawaii Telescope (CFHT) equipped with the MegaCam camera (Figure 1, right) in the $g$-band with a total exposure of 0.7h. In both images -- the Milankovi\'c  and the CFHT images -- the shell structure is prominent and the stellar stream stretching to the north-east is clearly visible.
 
The astrometric solution was found using the Astrometry code (Lang et al. 2010). The data reduction was done using the Milankovi\'c pipeline (M\"uller et al. 2019)  following standard procedures: bias, dark, and flat field corrections. The creation of the super sky flat was achieved trough stacking individual frames using their native (image) scale. Simply, they were all median combined ignoring their astrometric solution. This super sky flat image was then normalized to the median value of each individual science frame and subtracted from them. Finally, the subtraction of the super sky flat was done on all the individual frames that were previously bias, dark, and flat field corrected. The last step was the co-addition of individual frames using IRAF's { \tt imcombine} task.

\begin{table}
\label{phot}
\begin{center}
\begin{tabular}{|c|c|c||c|c|c|}
\hline
Time &$B$-limit & $L$-limit & Time & $V$-limit & $I$-limit \\

[Hours] & [mag/arcsec$^2$] &[mag/arcsec$^2$] & [Hours] & [mag/arcsec$^2$] &[mag/arcsec$^2$] \\ 
\hline
0.25 & 25.15 $\pm$ 0.02 & 27.27 $\pm$ 0.01 & 0.15 &  26.12 $\pm$ 0.01 &  26.54 $\pm$ 0.03 \\
0.75 & 25.89 $\pm$ 0.03 & 27.86 $\pm$ 0.01 & 0.45 &  26.79 $\pm$ 0.01 &  27.33 $\pm$ 0.02 \\
1.25 & 26.22 $\pm$ 0.02 & -- & 0.75 &  27.11 $\pm$ 0.02 &  27.65 $\pm$ 0.02 \\
1.75 & 26.42 $\pm$ 0.02 & 28.27 $\pm$ 0.02 & 1.05 &  27.22 $\pm$ 0.06 &  27.80 $\pm$ 0.03 \\
2.25 & 26.58 $\pm$ 0.04 & 28.35 $\pm$ 0.03 & 1.35 &  27.39 $\pm$ 0.01 &  27.89 $\pm$ 0.02 \\
2.75 & 26.68 $\pm$ 0.03 & 28.48 $\pm$ 0.03 & 1.65 &  27.35 $\pm$ 0.02 &  27.91 $\pm$ 0.02 \\
3.25 & 26.77 $\pm$ 0.05 &-- & 1.95 &  27.40 $\pm$ 0.01 &  27.94 $\pm$ 0.04 \\
3.75 & 26.85 $\pm$ 0.03 & -- & 2.25 &  27.54 $\pm$ 0.03 &  27.96 $\pm$ 0.04 \\
\hline
\end{tabular}
\caption{Surface brightness 3-sigma limits with the errors in the $B$-, $V$-, $I$- and $L$-band depending on the integral exposure time (in hours).}
\end{center}
\end{table}
\begin{figure}
\centering
\includegraphics[width=0.45\textwidth]{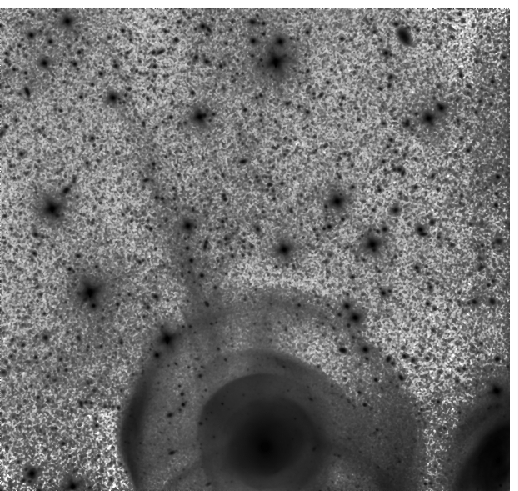}
\includegraphics[width=0.45\textwidth]{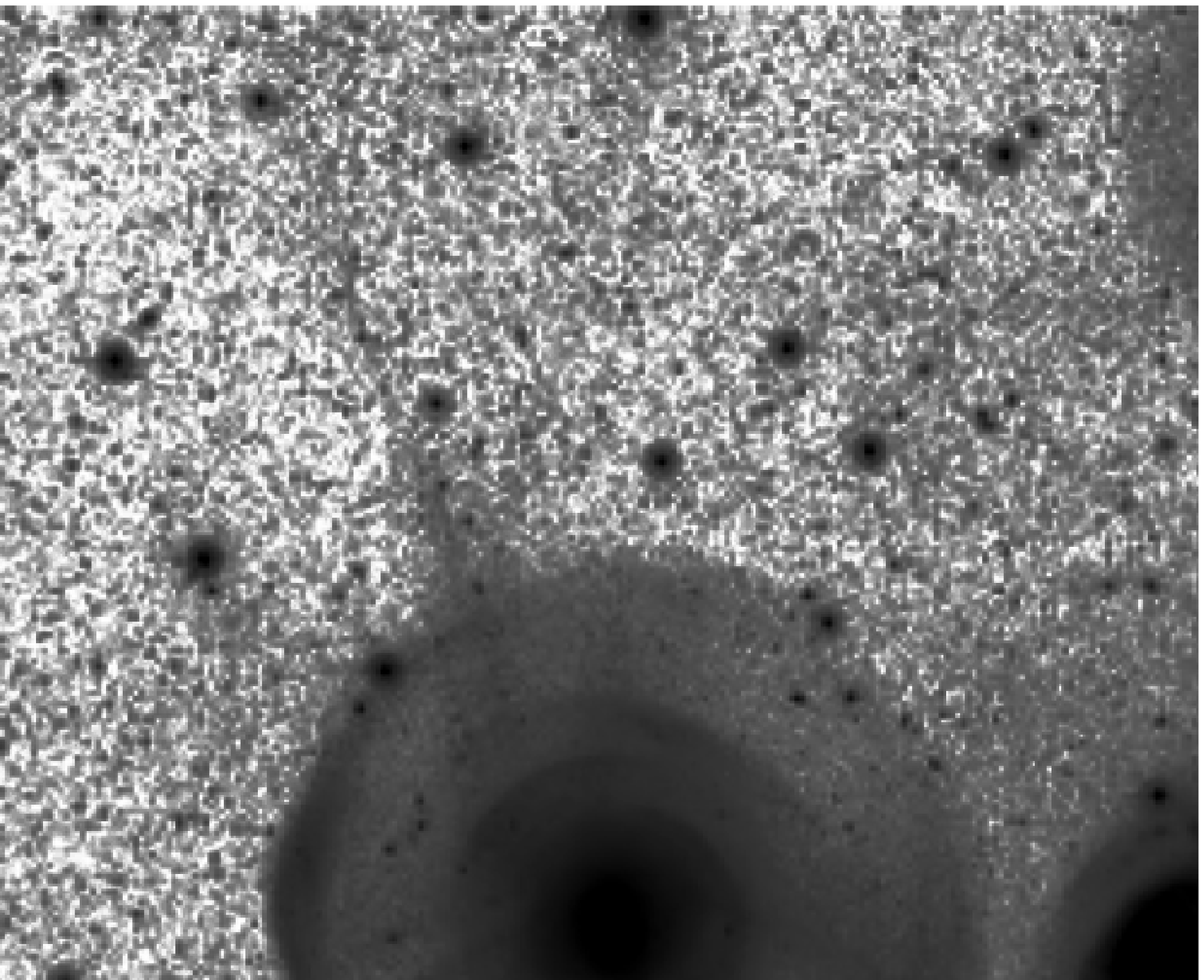}
\caption{NGC 474: ({\it left}) $L$-band image from the Milankovi\'c telescope and ({\it right}) $g$-band image from the CFHT/MegaCam (Duc et al.~2015).}
\end{figure}
 
\subsection{NGC~467 galaxy: $B$-, $V$- and $I$-band} 

We have observed galaxy NGC~467 on September 28$^{\rm th}$ and 29$^{\rm th}$ 2019. The same equipment was used as with NGC~474. We have employed three filters: $B$, $V$ and $I$. The individual exposure times were 300s for the $B$-band, and 180s for $V$- and $I$-band. In total, 47 images were taken in each band equaling to 3.9h in the $B$-band and 2.35h in the $V$- and $I$-bands. Dithering was $\sim$ 200 pix ($\sim$ 1 arcmin) and the seeing 1.3 arcsec. The final co-add image has the size of 16.3 $\times$ 16.6 arcsec$^2$. The same procedure as before was applied: (1) astrometric solution was obtained with Astrometry code and (2) the Milankovi\'c pipeline was applied to reduce the raw data. 
\section{SURFACE BRIGHTNESS LIMIT}

 Magnitudes were measured with the standard IRAF  package {\tt daophot} in the final co-add images of both galaxies in all filters. The photometric calibration was done using the linear regression formula:
 $m_f = a_0 + a_1 * m_{\rm cal}$,  where the subscript $f$ stands for filters used ($B$,$V$,$I$, and $L$), $a_0$, $a_1$ are the intercept and slope, and $m_{\rm cal}$ is the standard star magnitude from the photometric catalog used for the calibration.
 
 We have calibrated the magnitudes using the TOPCAT software (Taylor 2005). The magnitude zeropoint $a_0$ and the corresponding slope $a_1$ inside the whole FOV of each galaxy were inferred from SDSS DR12 photometric catalog using $g$-band magnitudes (Alam et al. 2016). They are: 32.7 $\pm$ 0.2 and 1.01 $\pm$ 0.02 ($L$-band), 29.50 $\pm$ 0.2 and 1.01 $\pm$ 0.02 ($B$-band), 30.80 $\pm$ 0.3 and 1.03 $\pm$ 0.03 ($V$-band), 31.6 $\pm$ 0.9 and 1.1 $\pm$ 0.01 ($I$-band). They are all expressed in the $g$-band. All stars brighter than the 22$^{\rm nd}$ magnitude and fainter than 14$^{\rm th}$ magnitude in the $g$-band were selected for measurements. 

Our main interest was to determine, apart from the final surface brightness limit, the limit reached throughout the observing run. To estimate how deep the images were in shorter exposures than the integral one, we have created lists of images with integral exposure time starting from 15 minutes up to 3.75 hours increasing with steps of 30 minutes. Lists of individual frames were created to correspond to these shorter exposures and each list was combined with IRAF's {\tt imcombine} task to make a stack. In each stack we measured the standard deviation inside a dozen boxes of 10 arcsec sides in empty regions (without objects). Standard deviations were averaged and their standard deviations were reported as the errors of the surface brightness limit converted to the $g$-band (Table 1). 

The results of the surface brightness limiting magnitudes in various bands were converted to the $g$-band using linear regression and compared to the commonly used sky surveys SDSS, BASS and DES (Figure 2). It appears that in about an hour and a half the depth of the SDSS has been reached with our observations in all the filters. During the same time the BASS and DES limiting magnitudes are reached with all the filters except for the $B$-filter. And in the particular case of the $L$-filter all the limits are exceeded in less than hour. However, using the $L$-filter in 7.2 hours we have reached the 3-sigma surface brightness limit of 28.4 $\pm$ 0.04 mag/arcsec$^2$ in the $g$-band (M\"uller, Vudragovi\'c \& B\'ilek 2019). This appears to be the limit of the $L$-band observations with the given strategy.
Here, this limit has been reached in less than 3 hours (Figure 2).
\begin{figure}
\centering
\includegraphics[width=8.cm]{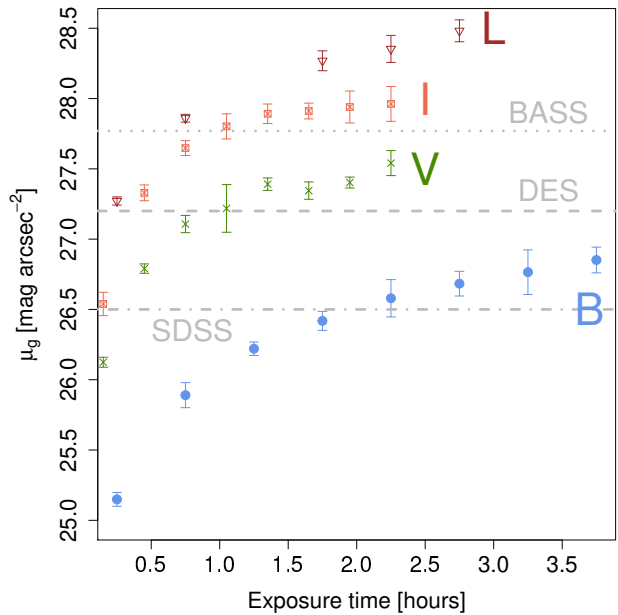}
\caption{Surface brightness limit (3-sigma) in the $B$-, $V$-, $I$- and $L$-band variation with exposure time compared to: SDSS, BASS and DES surveys (green dashed lines).}
\end{figure}


\section{DISCUSSION}

We have measured the limiting surface brightness in the $B$-, $V$-, $I$- and $L$-band using observations of the two nearby galaxies imaged with the Milankovi\'c telescope, equipped with IkonL CCD camera. The results can be used as a guideline for the exposure time needed for a given target surface brightness.

Comparing to other sky surveys (SDSS, BASS and DES), we have shown that in $\sim$ 1.5h we can reach greater depth. Moreover, our $L$-band observations exceed the limits of these surveys after only 45 minutes of on-target time. 
However, with the currently used imaging strategy, even after 7.2h integral exposure time, we couldn't breach the 28.4 mag/arcsec$^2$ $g$-band 3-$\sigma$ limit. 
A larger dithering pattern, including the rotation of the telescope, may overcome this issue. Despite this, our observations show that the Milankovi\'c telescope is capable of studying the low-surface brightness Universe and can be used to follow-up structures identified in other surveys, as well as to hunt for hitherto undetected features.


\smallskip
\centerline{\bf Acknowledgements}
\smallskip
We acknowledge the financial support of the Ministry of Education, Science and Technological Development of the Republic of Serbia (MESTDRS) through the contract No 451-03-68/2020-14/200002 and the financial support by the European Commission through project BELISSIMA (BELgrade  Initiative  for Space  Science,  Instrumentation  and  Modelling  in Astrophysics,  call  FP7-REGPOT-2010-5,  contract  No.  256772), which was used to procure the Milankovi\'c 1.4 meter telescope with the support from the MESTDRS.
O.M. is grateful to the Swiss National Science Foundation for financial support.
M.B. acknowledges the financial support by {\it Cercle Gutenberg}. The authors are grateful to Pierre-Alain Duc for providing the image of NGC~474 galaxy shown in Figure 1. We thank the technical operators at the Astronomical Station  Vidojevica (ASV), Miodrag Sekuli\'c and Petar Kosti\'c for their excellent work.
\references

Alam, S.,  Albareti, F.~D., Allende P.~C. et al. : 2016,\journal{VizieR Online Data Catalog}, V/147.

Bertin, E. \& Arnouts, S. : 1996, \journal{ 
Astron. and Astroph. Supplement S.}, \vol{117}, 393.

Duc, P.~A., Cuillandre, J.-C., Karabal, E. et al. :  2015, \journal{Mon. Not. R. Astron. Soc.}, \vol{446}, 120

Dey, A., Schlegel, David J., Lang, D. et al. : 2019, \journal{Astron. J.}, \vol{157}, 168

B\'ilek, M., Duc, P.-A., Cuillandre, J.-C. et al. : 2020, \journal{Mon. Not. R. Astron. Soc.}, \vol{498}, 2138

Bertin, E., Mellier, Y., Radovich, M. et al. :  2002, \journal{Astronomical Data Analysis Software and Systems}, \vol{281}, 228.

Bullock, J.~S. \& Johnston, K. V. : 2005, \journal{ Astrophy. J.}, \vol{635}, 931.

Cooper, A.~P., Cole, S., Frenk, C.~S. et al. 2010, \journal{Mon. Not. R. Astron. Soc.}, \vol{406}, 744

Frenk, C.~S. \& White, S.~D.~M. : 2012, \journal{ Annalen der Physik}, \vol{524}, 507

Lang, D., Hogg, D.~W., Jester, S. et al. : 2009, \journal{Astron. J.}, \vol{137}, 4400.


Rich, M.~R., Brosch N., Bullock, J. et al. : 2020, DOI: 10.1093/mnras/staa678. 

M\"uller, O., Vudragović, A. \& B\'ilek, M. : 2019, \journal{Astron. \& Astroph.}, \vol{632}, L13.  

Taylor, M.~B. : 2005, \journal{Astronomical Society of the Pacific Conference Series}, \vol{347}, 29pp.

van Dokkum, P.~G., Abraham, R. \& Merritt A. : 2014,
\journal{Astrophys. J.}, \vol{782}, 2.

York, D.~G., Adelman, J., Anderson, John E.~J., et al. : 2000, \journal{Astron. J.}, \vol{120}, 1579.

Zou, H., Zhou, X., Fan, X. et al. : 2017, \journal{ Publications of the Astronomical Society of the Pacific}, \vol{129:064101}, 9pp.


\endreferences

\end{document}